\documentclass{aa}
\usepackage{graphicx}
\begin{document}
   \title{NGC~3628: Ejection Activity Associated with Quasars}

   \subtitle{}

   \author{H. Arp\inst{1}
          \and
          E. M. Burbidge\inst{2}
          \and 
          Y. Chu\inst{3}
          \and
          E. Flesch\inst{4}
          \and
          F. Patat\inst{5}
          \and
          G. Rupprecht\inst{5}
          }

   \offprints{H. Arp; e-mail: arp@mpa-garching.mpg.de}

   \institute{Max-Planck-Institut f\"ur Astrophysik, 85741 Garching, Germany
      \and
       Center for Astrophysics and Space Sciences, University of California, 
       San Diego, La Jolla, CA 92093-0424    
       \and 
       Center for Astrophysics, University of Science and Technology, 
       Hefei, Anhui, 230026, China 
       \and 
       P.O. Box 12520, Wellington, New Zealand
       \and
       European Southern Observatory, 85748 Garching, Germany
             }

   \date{Received ??-??-2002; accepted ??-??-2002}

\abstract{
NGC~3628 is a well-studied starburst/low level AGN galaxy in the Leo 
Triplet noted for its extensive outgassed plumes of neutral hydrogen.  
QSOs are shown to be concentrated around NGC~3628 and aligned with the HI 
plumes. The closest high redshift quasar has $z$=2.15 and {\it is at the tip 
of an X--ray filament emerging along the minor axis HI plume.} Location at 
this point has an accidental probability of $\sim 2\times10^{-4}$. In 
addition {\it a coincident chain of optical objects coming out along the 
minor axis ends on this quasar.} 

More recent measures on a pair of strong X--ray sources situated at 3.2 and 
5.4 arcmin  on either side of NGC~3628 along its minor axis, reveal that 
they have nearly identical redshifts of $z$=0.995 and 0.981. {\it The closer 
quasar lies directly in the same X--ray filament which extends from the 
nucleus out 4.1 arcmin to end on the quasar of $z$=2.15.}
 
The chain of objects SW along the minor axis of NGC~3628 has been imaged in 
four colors with the VLT. Images and spectra of individual objects within 
the filament are reported. It is suggested that material in various physical 
states and differing intrinsic redshifts is ejected out along the minor axis 
of this active, disturbed galaxy.
 

}

   \maketitle
%

\section{Introduction}

NGC~3628 ($z$=0.0028) is a nearby, edge-on Sbc peculiar galaxy which is 
undergoing major internal dynamic activity that is, however, shrouded from 
our view by a prominent dust lane. See Atlas of Peculiar Galaxies No. 317 
(Arp 1966). For a comprehensive summary of the galaxy parameters 
and current observational status, see Cole, Mundell and Pedlar (1998).  
For a discussion of a possible AGN within NGC~3628, see Yaqoob et al. (1995).

A prominent feature of NGC~3628 is the long HI plumes being outgassed from 
the galaxy in two directions.  The major plume to the ENE was best imaged by 
Kormendy and Bahcall (1974) as a long, straight optical jet. Later it was  
observed in HI with the Arecibo telescope by Haynes, Giovanelli and Roberts 
(1979) who also mapped a weaker plume towards the south along the minor 
axis. The usual explanation for the plume morphology is that it is due to a 
tidal encounter between NGC~3628 and the nearby similar-sized spiral 
NGC~3627, although the velocity profile and substructure of the plume caused 
Haynes et al. to comment that "...the observational data somewhat strain the 
model parameters."  It could also have been noted that comparable HI 
extensions were not drawn out of adjoining galaxies with which NGC~3628 was 
supposed to have interacted. 

\begin{figure*}
\centering
\includegraphics[angle=0,width=17cm]{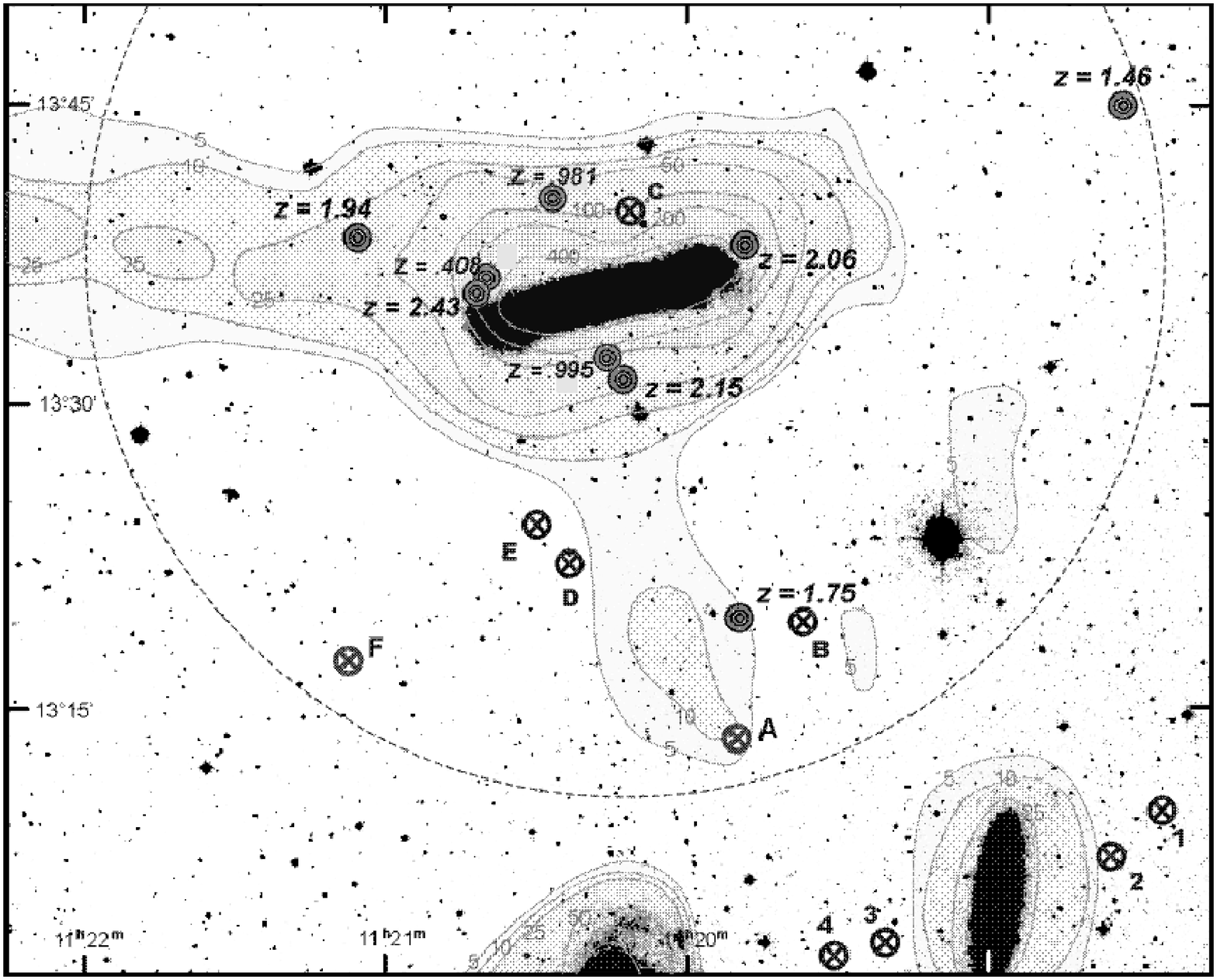}
\caption{Isophotal contours of neutral hydrogen (HI) are shown coming from 
the starburst$/$AGN galaxy NGC~3628 (from Haynes, Giovanelli and Roberts 
(1979).  Catalogued quasars from Weedman (1985) and Dahlem (1996) are 
annotated with their redshift values. The search field of Weedman is bounded 
by the dashed line.  Objects marked with a circled X are probable quasars 
plus F, a possible quasar (see Tab.~\ref{tab:wee} for detailed data). }
\label{fig:hydro}
\end{figure*}

Haynes et al.'s definitive mapping of the plume morphology is replicated 
here in Fig.~\ref{fig:hydro}, minus the complete extension to the ENE, and 
without the 3K km s$^{-1}$ contour about which Haynes et al. counselled 
caution. The velocity profile of the strongest plume is that of steady flow 
velocity away from the galaxy out to beyond the left edge of 
Fig.~\ref{fig:hydro}. 

This extended hydrogen is originating from the central regions of NGC~3628, 
either from starburst activity or from a dust-enshrouded AGN.  Fabbiano, 
Heckman and Keel (1990) concluded that the X-ray observations 
demonstrated "... collimated outflow from a starburst nucleus...", and Irwin 
and Sofue (1996) found expanding molecular shells of CO emanating from the 
nucleus.  NGC~3628 is known to have a strong X-ray source in its core, and, 
unusually, a second strong X-ray source toward the east end of its disk.  
Previous work by Radecke (1997), Arp (1997, 1998) and others has shown a 7.4 
sigma association of quasars with Seyfert galaxies, which we have attributed 
to ejection from their active nuclei.  Since NGC~3628 is a prominent nearby 
galaxy which is so clearly seen to be expelling material from an 
active nucleus, we looked to see if there are quasars close to this galaxy.  
As can be seen in Fig.~\ref{fig:hydro}, there are indeed known quasars in 
the near vicinity of NGC~3628, even in near proximity to its active 
disk.  In this paper we will discuss evidence that these quasars are being 
ejected from NGC~3628 along with the HI gas and X-ray material.

\begin{table*}
\caption[]{Objects in Fig.~\ref{fig:hydro}.}
\label{tab:wee}
\centering
\tabcolsep 5pt
\begin{tabular}{llcccccl}
\hline
Obj    & Survey ID & Prob. QSO & R.A. (J2000) & Decl & E & O & Notes \\
\hline
z=1.46 & Wee 47    & QSO   & 11 18 30.3 & 13 45 01 & 18.5 & 18.9 & No X-ray\\
z=2.06 & Wee 48    & QSO   & 11 19 46.9 & 13 37 59 & 19.2 & 20.0 & 2RXP 6 cts/hr\\
z=1.75 & Wee 50    & QSO   & 11 19 48.2 & 13 19 38 & 19.3 & 19.8 & No X-ray \\
z=2.15 & Wee 51    & QSO   & 11 20 11.9 & 13 31 23 & 19.4 & 19.6 & 1RXH 5 cts/hr \\
z=0.995& 1WGAJ1120.2+1332&QSO&11 20 14.7& 13 32 28 & 19.5 & 20.7 & 1RXH 3 cts/hr \\
z=0.981& 1WGAJ1120.4+1340&QSO&11 20 26.2& 13 40 24 & 18.6 & 19.8 & 1RXH 9 cts/hr \\
z=0.408& 1WGAJ1120.6+1336&QSO&11 20 39.9& 13 36 20 & 19.1 & 20.1 & 1RXH 3 cts/hr \\
z=2.43&  Wee 52    & QSO   & 11 20 41.6 & 13 35 51 & 19.5 & 20.3 & 1RXH 3 cts/hr \\
z=1.94&  Wee 55    & QSO   & 11 21 06.1 & 13 38 25 & 18.0 & 18.7 & 2RXP 22 cts/hr\\
      &            &       &            &          &      &      & \\
A     &  Wee 49    &Probable&11 19 48.3 & 13 13 30 & 19.9 & 20.6 & Two EM lines \\
B     &  2RXPJ111935.0+131921&90\%&11 19 35.1&13 19 24& 18.6&19.6& 7 cts/hr\\
C     &  1RXHJ112010.3+133939&74\%&11 20 10.5&13 39 34& 18.8&21.0& 16 cts/hr\\
D     &  2RXPJ112022.6+132212&97\%&11 20 22.8&13 21 58& 19.8&20.4& 9 cts/hr\\
E     &  2RXPJ112028.8+132416&85\%&11 20 29.4&13 23 57& 19.7&20.4& 14 cts/hr\\
F     &  Wee 56    &Possible&11 21 09.7 & 13 17 53 & 19.9 & 20.9 & One EM line\\
      &            &       &            &          &      &      & \\
1     &  1RXHJ111821.9+130957&71\%&11 18 21.7&13 09 57& - & 21.0 & 2 cts/hr\\
2     &  1RXHJ111832.6+130732&100\%&11 18 32.5&13 07 32&19.1&20.0& 6 cts/hr\\
3     &  1RXHJ111919.1+130315&85\%&11 19 18.9&13 03 17& 19.8&20.6& 5 cts/hr\\
4     &  1RXHJ111928.8+130249&99\%&11 19 28.4&13 02 51& 17.6&18.4& 12 cts/hr\\ 
\hline
\end{tabular}
\end{table*}


\section{Quasars and candidates near NGC~3628}

Dahlem et al. (1996) have listed many X-ray emitting sources in 
the ROSAT-detected hot gaseous halo of NGC~3628.  They found that this 
density of sources was over twice that of the wider background (=1.5 sigma 
deviation), but made no firm statistical statement because only small numbers 
were involved.  Dahlem et al. note that the emission properties of these 
sources point to their identification as either AGN (QSOs) or X-ray binaries.
Since it is unlikely that powerful binary sources are present in the halo it 
is much more probable that these sources are more closely related to AGNs. 
In the standard model these QSOs would be assumed to be  background objects, 
their greater density near NGC~3628 notwithstanding.  In the alternative 
model advanced by Arp (1987), G.R. Burbidge and others, most of these QSOs 
are associated with NGC~3628, as is the case for other galaxies with 
physically associated quasars, see e.g., E.M. Burbidge (1999).

\begin{figure}
\resizebox{\hsize}{!}{\includegraphics{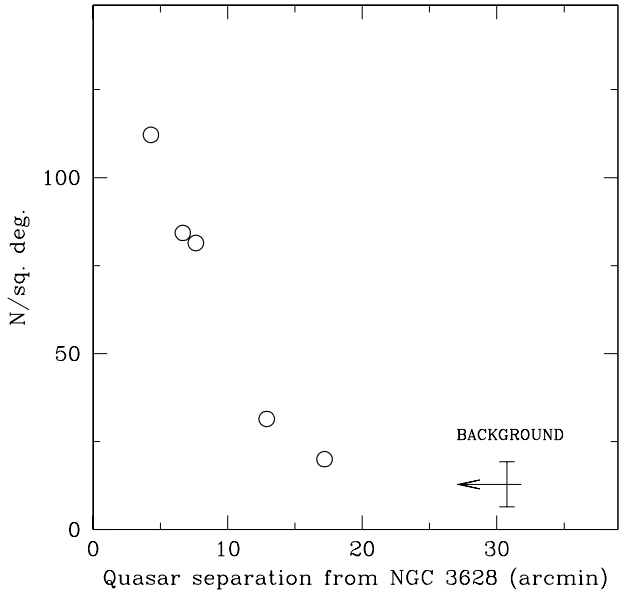}}
\centering
\caption{The density of Weedman catalogued quasars at the distance of each
quasar from NGC~3628 is plotted. The arrow marks the upper limit to average
background density of 1.75 $\leq z \leq$ 2.5 quasars. The area blocked by
NGC~3628 has been excluded.}
\label{fig:counts}
\end{figure}

Fig.~\ref{fig:hydro} shows 15 confirmed or probable QSOs in the surveyed 
area, of which 10 are within the galaxy's plume contours. The average 
background density which Weedman (1985) found for his 20 CFHT grism fields 
was 10/sq. deg. for $2.0 \leq z <  2.5$ to m$_{4500}$ = 21 mag. 
For $1.75 \leq z < 2.5$ the density rises to 13.9/sq deg +8/-6 and this 
background density is plotted in Fig.~\ref{fig:counts}. 

It is seen that the closest Weedman quasars to NGC~3628 reach a density of 
about 100/sq. deg. The chance of finding the two closest quasars at 19.7 and 
19.9 mag. is only 0.002 (using Poisson statistics). Of the three closest, the 
QSO with $z$=2.43 is 21.2 mag., probably fainter due to absorption by the 
galaxy, but still making the probability of chance concentration even 
smaller. This probability is an upper limit because the background counts 
are over estimated. 
This is due to the fact that, as we have seen, the Weedman survey 
inadvertently pointed near some very active galaxies and then only used 
counts from plates which showed the most quasars below m$_{4500}$ = 20 mag. 

Tab.~\ref{tab:wee} summarizes what is known about the quasars and quasar candidates 
depicted in Fig.~\ref{fig:hydro}. In Tab.~\ref{tab:wee} confidence values that these 
objects are QSOs are displayed as percentages.  These are taken from the 
whole--sky X-ray/radio/optical overlays catalogue by Flesch, which is 
accessible at ftp://quasars.org/quasars (paper in preparation).
   
\begin{figure*}
\centering
\includegraphics[width=17cm]{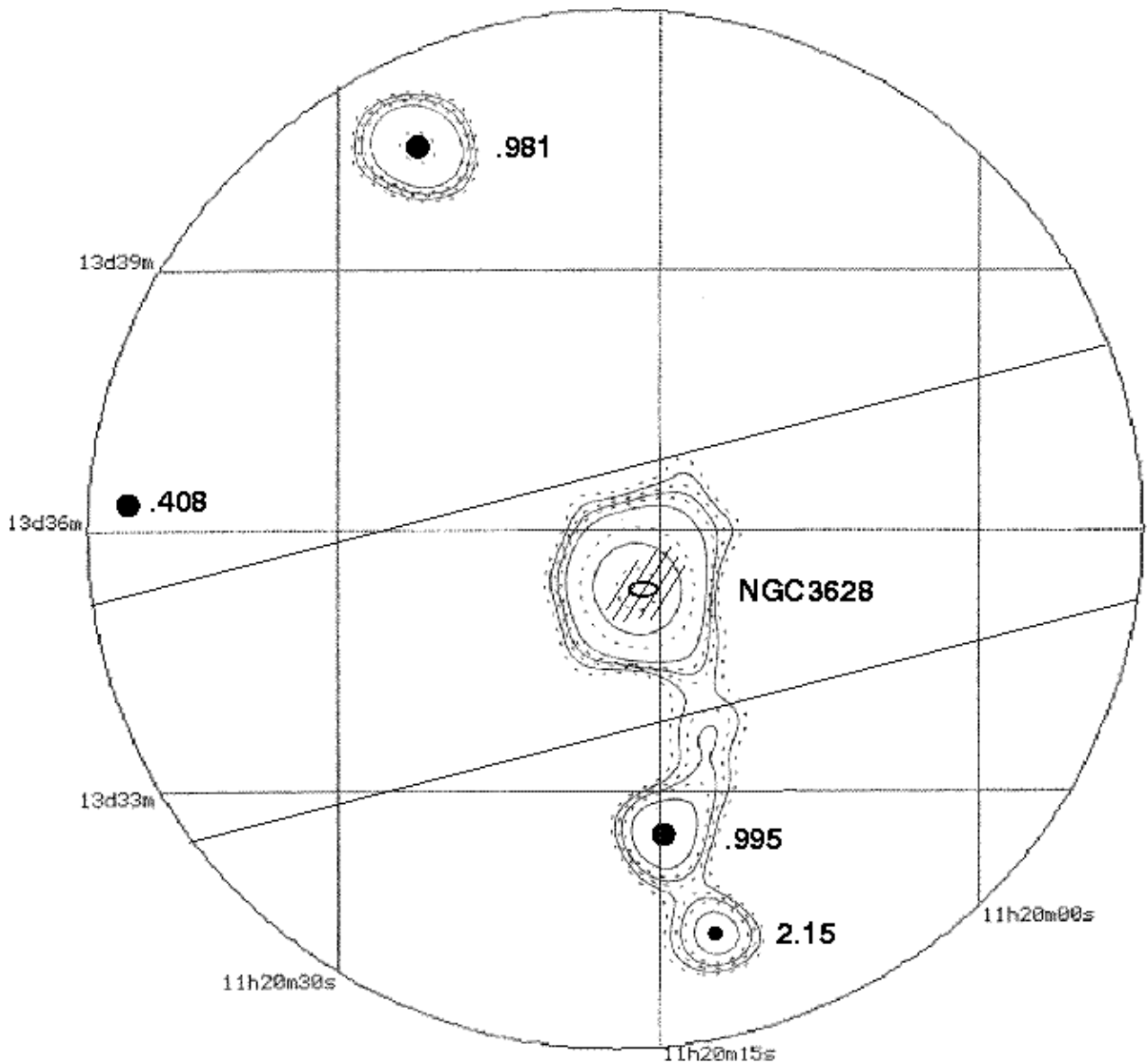}
\caption{The broad band (0.1 to 2.4 keV) ROSAT map of NGC~3628 smoothed and 
contoured from PSPC photon event files. It is quite similar to the map 
presented by  Dahlem et al. (1996).  The optical position of the X-ray 
quasars in the field, however, have now been indicated as filled circles 
with their redshifts labelled.  }
\label{fig:rosat}
\end{figure*}

Briefly, the method is to first calculate likelihoods of X-ray-optical
association for each cross-category of X-ray-optical offset distance,
optical PSF, and O-E color, by comparing the found density of such optical
objects to the whole-sky background density of such objects, using a
650,000,000-object merge of the APM and USNO optical catalogues,
subcategorized by local sky densities to minimise local effects. These
likelihoods are then used to realign the astrometry of the ROSAT X-ray
fields with the optical background, then the likelihoods are recalculated,
and this process is iterated to stability.  Having thus obtained best
X-ray-optical positional fits, the most likely optical object is determined
for each X-ray detection, and where there is a good optical candidate, the
probability that it is a QSO is derived from the numerical prevalence of
known QSOs/galaxies/stars having that object's X-ray-optical positional
offset, PSF, O-E color, and X-ray-to-optical flux ratio, decreased by the
calculated probability that the X--ray association is false.  A fuller
accounting is provided by the readme which accompanies the catalogue at the
web address.

Tab.~\ref{tab:wee} also lists four X-ray quasar candidates around NGC~3623. 
Fig.~\ref{fig:hydro} shows that NGC~3623, an adjoining bright member of the 
Leo Triplet, has these four probable quasars aligned across its nucleus, two 
on either side. Redshifts of these candidates would furnish additional 
confirmation of ejection of quasars from these Leo Group galaxies.    
     

\section{Quasars in the galactic disk of NGC~3628}

In Fig.~\ref{fig:hydro} the $z$=2.43 quasar is the X-ray source Dahlem \#16. 
Only 37 arcsec away is the X-ray source Dahlem \#15 which we have identified 
with a blue stellar object (BSO) of similar apparent magnitude. RIXOS 
(Strickland et al. 2000) measures reported $z$=0.408 for the second component 
of this apparent double quasar. 
Optical identification of these objects was enabled from the coordinates of 
Read et al. (1997) and interchange of the published declinations of Dahlem 
\#15 and \#16 (M. Dahlem, private communication).

The quasars at both  the east and west end of the disk appear to be
associated with strong disturbances in NGC~3628 at these points. Luminous
features and dust features point in the general direction of the quasars
and there are hints of filaments and perturbations which, when explored 
with deeper, higher resolution images from larger telescopes, may link 
these quasars with the general eruption of material in these regions.


\section{Quasars in the HI Plumes}

Fig.~\ref{fig:hydro} shows alignment between the HI plumes of NGC~3628 and 
nearby quasars.  
Quasars with $z$=1.94, 2.43 and 0.408 lie at the base of the main ENE plume, 
coincident with the start of the optical jet imaged by Kormendy and Bahcall 
(1974). Two more, with $z$=2.06 and 1.46, align in roughly the opposite 
direction. Four more lie in the southern plume, $z$=0.995, 2.15. 1.75 and 
candidate A (Wee 49). The latter two lie along a thickening of the plume. 
The probable quasar Wee 49 lies, very interestingly, right at the tip of the 
southern HI plume.  Thus these quasars are not only aligned with the plumes, 
but positioned along contour nodes.  This is strongly indicative of physical 
association, and implies that these quasars and HI plumes have come out of 
NGC~3628 in the same physical process.


\section{X-ray Ejection from the Nucleus of NGC~3628}

As referenced previously, the first X-ray observations established collimated 
outflow along the minor axis of NGC~3628.  The later ROSAT observations 
confirmed this result and established narrow filaments and point sources 
extending outward from the nucleus (Dahlem et al, 1996; Read, Ponman and 
Strickland 1997).  Fig.~\ref{fig:rosat} here, the best resolution PSPC X-ray 
map, shows a narrow jet/filament coming from the bright X-ray nucleus 
continuously out to end at the $z$=2.15 quasar. 

The X-ray contours shown in Fig.~\ref{fig:rosat} were processed from ROSAT 
photon event files by Arp. The resultant X-ray maps in both broad band 
(shown in Fig.~\ref{fig:rosat}) 
and hard band show essentially the same connection from the nucleus of 
NGC~3628 to the $z$=2.15 quasar that the maps of Dahlem et al. (1996) show.

If we adopt the last X-ray source as the end of the filament, then the quasar
falls essentially at its tip. The accuracy of the superposition is obtained 
by taking the X-ray position of Dahlem \#7 and differencing it with that of 
the APM position of the quasar. That yields a displacement of $\sim$20 arcsec 
and a probability of accidental superposition of $\sim10^{-3}$. However, 
identifying optically the stronger Dahlem sources gives $\sim$12 arcsec
systematic correction for the X-ray positions. That yields a coincidence of 
8 arcsec, about the accuracy of PSPC identifications, and a probability of 
accidental coincidence of $2\times10^{-4}$.    

This quasar, as marked in Fig.~\ref{fig:rosat}, is an X-ray source and the 
next source in toward the nucleus also is a point X-ray source which has 
been measured by RIXOS at $z$=0.995.


\begin{figure*}
\centering
\includegraphics[width=15cm]{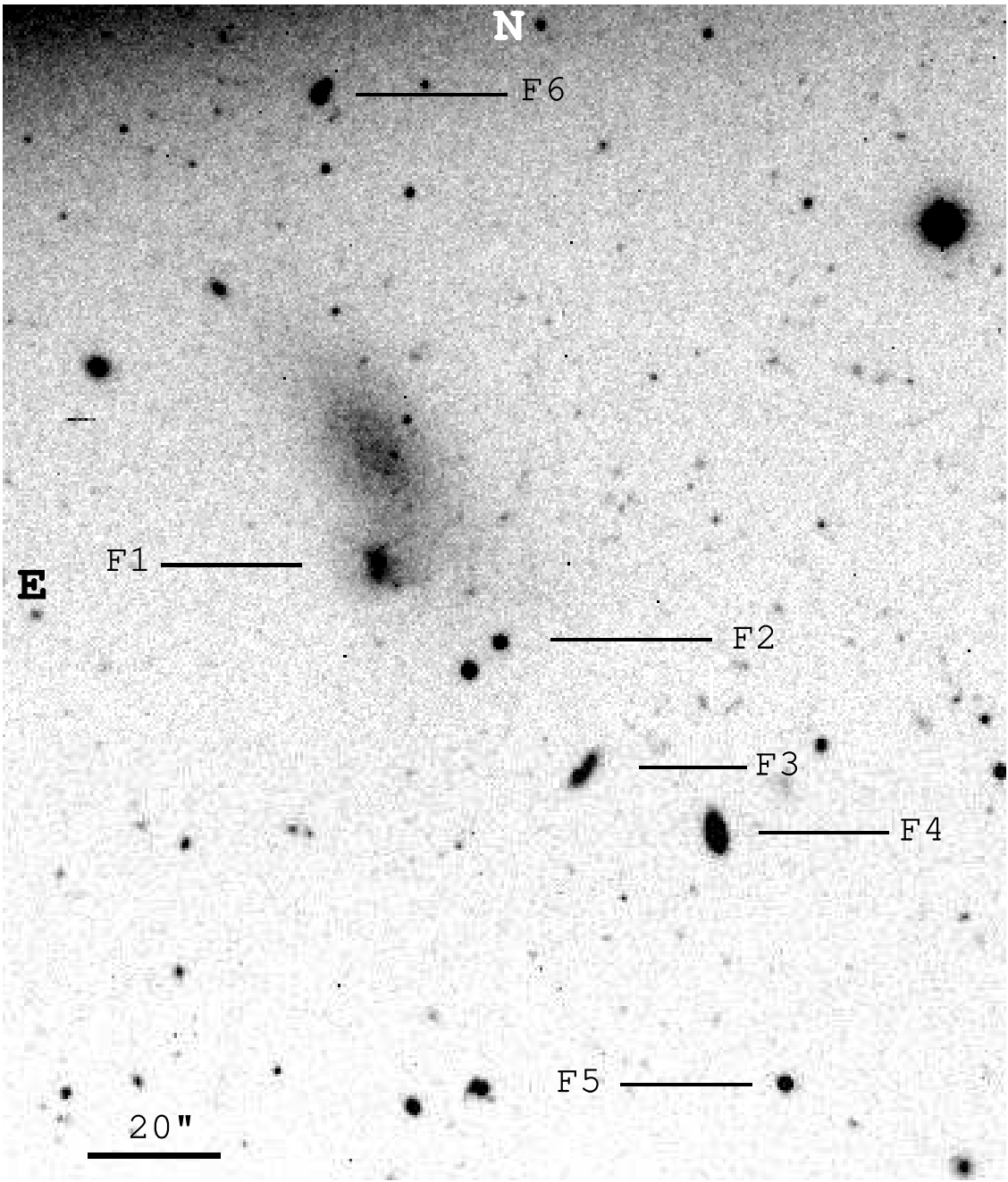}
\caption{$R$ band exposure with VLT-FORS2 of the chain of objects coming
SSW along the minor axis of NGC~3628. Quasars for Fig.~\ref{fig:rosat} map
and other optical objects in filament are identified as F1--F6. Available
spectral data is summarized in Tab.~\ref{tab:objects}. See ESO Messenger,
March 2002 for a color picture of this SW region of NGC~3628.}
\label{fig:fors}
\end{figure*}

\section{The Pair}

At a slightly greater distance on the other side of the NGC~3628 nucleus, 
along the N minor axis, is a brighter X-ray source which is identified with 
an E = 18.6 mag., blue stellar object (BSO), shown as object C in 
Fig.~\ref{fig:hydro}.  This was observed to be a quasar of $z$=0.984 by 
E.M Burbidge in 1999:

\vspace{5mm}
\tabcolsep 20pt
\begin{tabular}{ccc}
\hline
 Line  & $\lambda_{obs}$ & z \\
\hline
 Mg II 2800 & 5565 & 0.988 \\
 CIII] 1909 & 3780 & 0.980 \\
\hline
\end{tabular}
\vspace{5mm}

Later RIXOS (Mason et al. 2000) reported for the same two lines a redshift 
of  $z$=0.977. We adopt $z$=0.981 as the mean of the two sets of measures. 
Together with the $z$=0.995 quasar discussed at the end of the previous 
section, {\it these two X-ray quasars form a pair aligned across the 
nucleus of NGC~3628} like so many other pairs which have now been reported 
across active galaxies (see Arp 1997; 1998).

Another striking aspect of this pair, however, is that they lie closely 
along the minor axis and very close to the nucleus of NGC~3628 (3.2 and 5.4 
arc min). They have very similar redshifts, when transformed to the rest 
frame of NGC~3628, $z$=0.98 and 0.99. These redshifts are closely matched -- 
a characteristic of many previous pairs of quasars across active galaxies -- 
and demonstrate how unlikely it is that they are unassociated background 
objects. It is also noteworthy that these two redshifts are very close to 
the Karlsson value of quantized redshift of $z$=0.96. The remaining quasars 
in this field tend to have redshifts near the preferred peaks (for recent 
analyses of periodicities in z see Burbidge and Napier 2001).

In summary we see that in the case of the nearby galaxy NGC~3628, not only 
an HI plume but narrow PSPC X-ray filaments emerge along the minor axis 
connecting the $z$=2.15 and $z$=0.995 quasars back to the active nucleus. In 
addition the quasar in the filament has a counterpart, closely matched in 
redshift, on the other side of NGC~3628. (The recent Chandra map of the inner 
2$\times$2 arcmin (Strickland et al. 2001) shows X-ray material leading 
back to the NNE in the direction of the $z$=0.981 quasar. It would be of 
considerable interest to see what the rest of the Chandra field showed with 
regard to the narrow filament 4.1 arcmin to the SSE in which the two quasars 
are embedded).
 

\begin{figure}
\resizebox{\hsize}{!}{\includegraphics{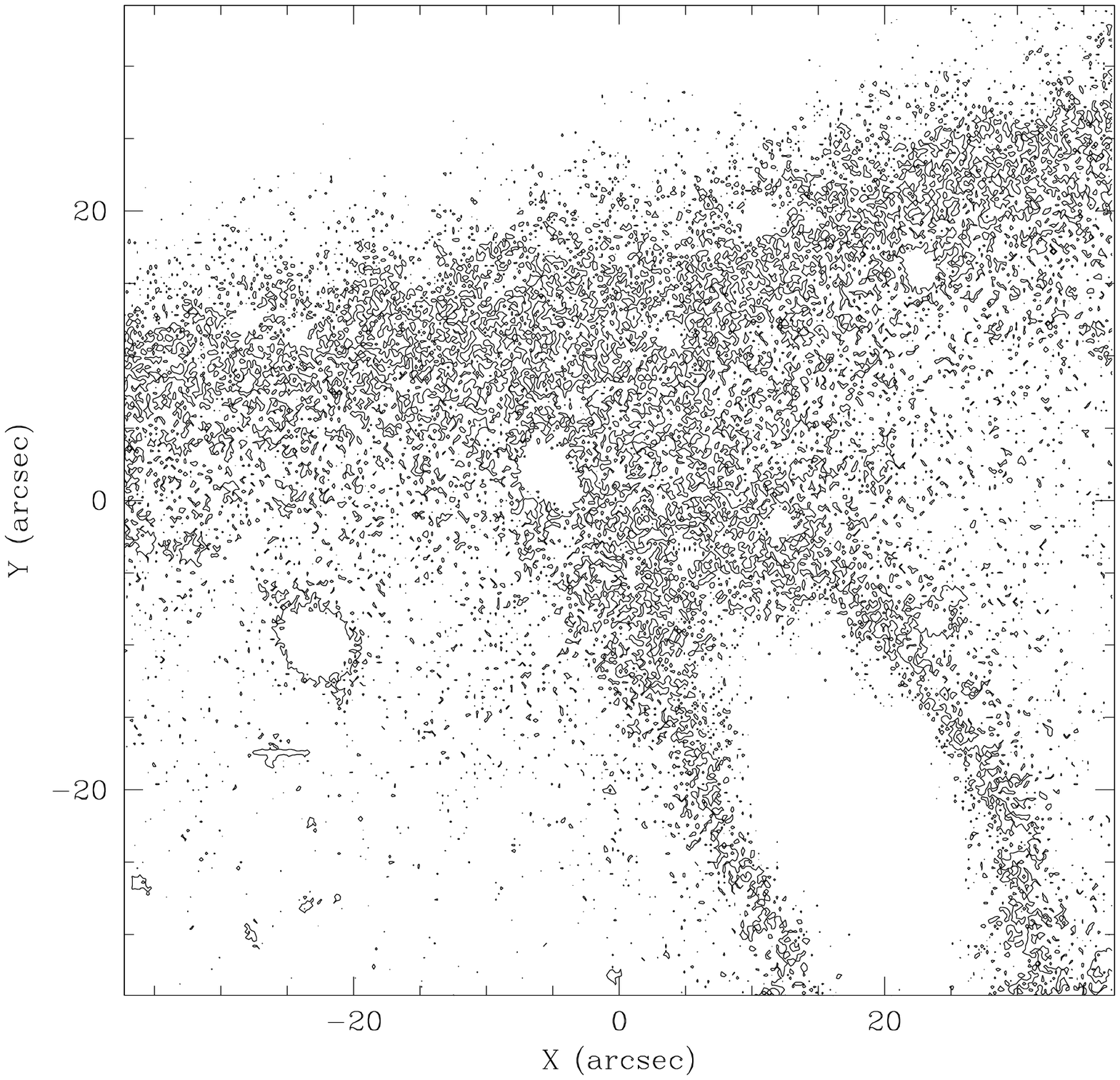}}
\centering
\caption{Low surface brightness object in chain of objects along the
SW minor axis of NGC~3628. From the $R$ image of the VLT, the lowest contour
here is 5 sigma above the sky background and suggests a connection with the
main body of NGC~3628.}
\label{fig:contour}
\end{figure}

\section{The Optical Objects Leading to the $z$=2.15 Quasar}

It is particularly interesting that a chain of optical objects coincides 
with the narrow X-ray filament. It is conspicuous on both red and blue 
Poss~II schmidt survey plates (Flesch and Arp 1999).  This feature has been 
imaged with the ESO VLT at Paranal, Chile and is shown in color in the ESO 
Messenger (March 2002). Figure 4 here shows a FORS2 R band image of this 
optical filament. The pictures reveal that the object nearest the main 
galaxy is nebulous, shaped like a comet with a blue condensation at its head. 
Next along the filament to the SW are stellar- appearing objects, one quite 
red and one quite blue. The blue one turns out to be the $z$=0.995 X-ray 
quasar discussed previously (F2 in Fig.~\ref{fig:fors}). Further along is a 
small, double, irregular galaxy or knot (F3) and then a larger, elongated 
and curved galaxy shape (F4). It has a narrow emission line spectrum of 
redshift $z$=0.153  (measured by Chu and Zhu in 1999 and Burbidge and Arp in 
2000). Finally we encounter the $z$=2.15 quasar (F5) which is at the end of 
the X-ray filament. The data is summarized in Tab.~\ref{tab:objects}.

\begin{table*}
\caption[]{Objects in Fig.~\ref{fig:fors}.}
\label{tab:objects}
\centerline{
\tabcolsep 15pt
\begin{tabular}{cccccl}
\hline
ID   &  B-V    & z     & R.A.(2\rlap{000)~}& Dec. & Notes\\
\hline
  F1 &  0.87   & 0.038 & 11 20 16.1& 13 32 39 & bright knot S \\
  F2 &  0.51   & 0.995 & 11 20 14.7& 13 32 27 & X-ray, BSO \\ 
  F3 &  0.66   &  -    & 11 20 13.9& 13 32 05 & double blue galaxy \\ 
  F4 &  0.86   & 0.153 & 11 20 12.5& 13 31 55 & emission line, pec. galaxy\\
  F5 &  0.20   & 2.15  & 11 20 11.9& 13 31 23 & obj. prism, Wee 51 \\  
  F6 &  0.57   &  -    & 11 20 17.8& 13 34 50 & very blue, compact, double?\\
\hline
\end{tabular}
}
\end{table*}

 There is perhaps a precedent for this in the optical filament in the giant 
radio galaxy CenA (NGC 5128).  There the optical filament coincides with the 
direction of both the radio jet and the X-ray jet further in the interior.  
The optical filament consists of young stars and HII region-like emission 
lines, similar to a star forming arm in a spiral galaxy (Blanco et al. 1975; 
Arp 1986).  We might then expect a low redshift emission line spectrum for 
the optical filament in NGC~3628.  But in NGC~3628 there appear mostly 
discrete objects. The  $z$=2.15 quasar is at the tip of the filament and 
there are two X-ray point sources in the filament as well as several blue 
and/or slightly extended optical objects. 
 
In general, the narrowness of the filament requires that whatever is ejected 
in the X-ray jet (and in the case of Cen A the coincident radio jet) must be 
quite small.  There would seem to be no candidates other than quasars, which 
are generally X-ray and radio sources and exhibit similar spectra to the 
active nuclei which are actually ejecting the material.  Figures 3 and 4 
could then represent a fortuitous moment when the quasar is just passing out 
beyond the filament. If this is the case we should be able to investigate 
the mechanics of the entrainment and excitation by obtaining further spectra.

Finally it should be remarked that with the obvious disruption of NGC~3628 
it is expected that ejected material would be interacting with it as it 
emerged. One result would be to slow down the ejection speeds depending on 
the comparative densities of the matter involved and their particular route 
exiting the galaxy. This would result in ejecta being observed generally 
closer to their galaxy of origin and possibly disturbing the relation 
between redshift and distance from the galaxy (i.e. the higher 
redshift objects being closer to the galaxy) which is observed in ejections 
with less interaction (Arp 1999, Fig.~3).


\section{The Low Surface Brightness Object}

The most interesting spectrum to obtain, however is that of the comet 
shaped nebula at the beginning of the line of objects. To this end E. M. 
Burbidge, H. Arp and V. Junkkarinen placed a 3 arc sec wide slit of the 3 
meter Lick telescope N-S across the nebula and its condensation. Night sky 
subtraction and processing revealed only one faint, narrow emission line 
but it was slightly extended and present on both 1800 sec red exposures. If 
the observed emission line in this nebulous object, F1, is H$\alpha$ its 
redshift is $z$=0.038. This could represent entrained material in the 
ejection from the interior regions of NGC~3628 at  about 11,000 km s$^{-1}$ 
projected, or material of low intrinsic redshift or a mixture of both. 
The two quasars, F2 and F5, would have to be predominantly intrinsic in 
redshift. F3 is a double irregular object whose redshift is required. The 
spectrum of F4 looks like a narrow line galaxy at $z$=0.153 and may be 
non-associated although it is contorted. F6 is very blue and compact and 
requires a spectrum.

The comet shaped object, F1, is quite unusual and apparently not in a 
state of dynamical equilibrium. Since it is only about 25 arcsec from the 
quasar F2 it raises the possibility of interaction, either radiative or 
dynamical. It shows incipient resolution into compact objects in the R 
image. The bright condensation appears blue but there nevertheless may be 
considerable absorption which perhaps accounts for its very weak spectrum. 
Perhaps it resembles most a dwarf galaxy or a very large HII region. It is 
interesting to note that it has been suggested in the past that dwarf 
galaxies can be formed in disturbing events from larger galaxies  
(Arp 1996; Gallagher et al. 2001).  

Fig.~\ref{fig:contour} is included here to show that the NE end of the low 
surface brightness object joins the main body of NGC~3628 in an apparently 
straight sided channel. The faint contour shown is 5 sigma above sky and 
suggests that the object is like a dwarf emerging from the central regions 
of NGC~3628. 
Perhaps the gas has been stripped out of a young, dwarf stellar assemblage. 
To explore this object further it would be important to get deeper, higher 
resolution spectra to ascertain the kind of stars which seem to be on the 
verge of resolution and the nature of the bright condensation at its S end.  


\section{Summary}

In a completely searched area around NGC~3628 the known quasars are 
concentrated to the position of the starburst/AGN galaxy. At least four 
quasars are situated close to  the edge of the galaxy disk.  Further out the 
galaxy has prominent plumes of hydrogen gas and the quasars are well-aligned 
with key points on the plume contours.  Along the minor axis of NGC~3628 
there are a pair of quasars on opposite sides of the galaxy, a configuration 
of which there are now many known examples. The redshifts of these two X-ray 
quasars at $z$=0.977 and $z$=0.995 only differ by 0.018. 

{\it Perhaps most striking of all, a narrow X-ray and optical filament 
along the SSW minor axis connects the nucleus of NGC~3628 directly to a 
third quasar of $z$=2.15.}

Preliminary spectroscopic investigation along the filament show a mixture of 
high and low redshift objects presumably reflecting a mixture of entrainment 
velocities of ejection and/or intrinsic redshifts.

We believe the improbability of finding quasars so close to NGC~3628, 
including two of them linked directly to the nucleus by an X-ray filament, 
combined with finding the galaxy to be so actively ejecting associated 
plumes of gas, optical and X-ray material in these directions is key 
confirmation of the previous evidence for ejection origin of quasars.  
A search for further quasars located within the solid angle of the bright 
disk of NGC~3628, spectroscopic identification of remaining quasar 
candidates in the field and further analysis of the optical filament should 
give insight into the physical mechanisms of their origin.


\begin{acknowledgements}
Part of the data was provided by the ESO/ST - ECF Science 
Archive Facility.  
\end{acknowledgements}

\end{document}